\documentclass[useAMS,usenatbib]{mn2e}
\usepackage{graphicx,rotate}

\def\gtsim{\mathrel{\hbox{\rlap{\hbox{\lower4pt\hbox{$\sim$}}}\hbox{$>$}}}}
\def\lesssim{\mathrel{\hbox{\rlap{\hbox{\lower4pt\hbox{$\sim$}}}\hbox{$<$}}}}
\def\Msunpyr{M$_{\odot}\,$yr$^{-1}$}
\def\Msun{M$_{\odot}$}

\def\cm{{\rm\thinspace cm}}

\def\K{{\rm\thinspace K}}
\def\keV{{\rm\thinspace keV}}
\def\km{{\rm\thinspace km}}
\def\kpc{{\rm\thinspace kpc}}

\def\Mpc{{\rm\thinspace Mpc}}
\def\Msun{\hbox{$\rm\thinspace M_{\odot}$}}

\def\ps{{\rm\thinspace s^{-1}}}

\def\yr{{\rm\thinspace yr}}

\def\kmps{\hbox{$\km\ps\,$}}
\def\kmpspMpc{\hbox{$\kmps\Mpc^{-1}\,$}}

\def\Msunpyr{\hbox{$\Msun\yr^{-1}\,$}}

\def\pscm{\hbox{$\cm^{-2}\,$}}

\def\pccm{\hbox{$\cm^{-3}\,$}}
\def\pscm{\hbox{$\cm^{-2}\,$}}

\title{\emph{Chandra} observations of Abell 2199%
}

\author[Johnstone, et al.]
       {R. M. Johnstone\thanks{E-mail: rmj@ast.cam.ac.uk}, S. W. Allen,
        A. C. Fabian and J. S. Sanders\\
        Institute of Astronomy, University of Cambridge, Madingley Road,
        Cambridge CB3 0HA}
\date{%
      Received }

\pagerange{\pageref{firstpage}--\pageref{lastpage}}
\pubyear{}

\begin{document}

\maketitle

\label{firstpage}

\begin{abstract}
\noindent
We present results from an analysis of two {\it Chandra} observations of the
rich, nearby galaxy cluster Abell 2199. We find evidence (having
corrected for projection effects) for
radial gradients in
temperature and metallicity in the X-ray emitting gas: the
temperature drops from $kT\sim4.2$ keV at R=200~kpc to 1.6~keV within
R=5~kpc of the centre.
The metallicity rises from $\sim 0.3$ solar at R=200~kpc to
$\sim 0.7$ solar at R=30~kpc before dropping to 0.3 solar within the
central 5~kpc. We find evidence for structure in the surface
brightness distribution associated with the central radio source
3C338. No evidence is found for the gas having
a large spread in temperature at any particular location despite the
cooling time being short ($<10^9$yr) within the central $\sim15$~kpc.
Heating and mass cooling rates are calculated for various assumptions
about the state of the gas.

\end{abstract}


\section{Introduction}
\label{intro}

Abell~2199 is a nearby rich cluster of galaxies, the central dominant
galaxy of which is the cD galaxy, NGC~6166.  The intracluster gas is
of intermediate temperature (4.7 keV; \citealt{EdgeStewart91}) and
gives rise to X-ray emission which is strongly peaked on the central
galaxy, leading to the suggestion of a strong cluster cooling flow,
e.g., \citet{Peresetal98}.

NGC~6166 hosts the radio source 3C338 which has been known for nearly
20 years to have a peculiar morphology \citep{Burnsetal83}.
\citet{GeOwen94} have shown that the radio emission from 3C338 suffers
considerable depolarization and Faraday rotation which is consistent
with the presence of a dense intracluster medium. \citet{OwenEilek98}
have compared the ROSAT HRI X-ray image with a 5GHz radio map and
found evidence for a strong interaction between the radio source and
the X-ray emitting gas (with the X-ray gas disrupting the radio
source).  They argue that the intracluster medium is being heated by
the radio source.
 \begin{figure}
\protect\resizebox{\columnwidth}{!}
{\includegraphics{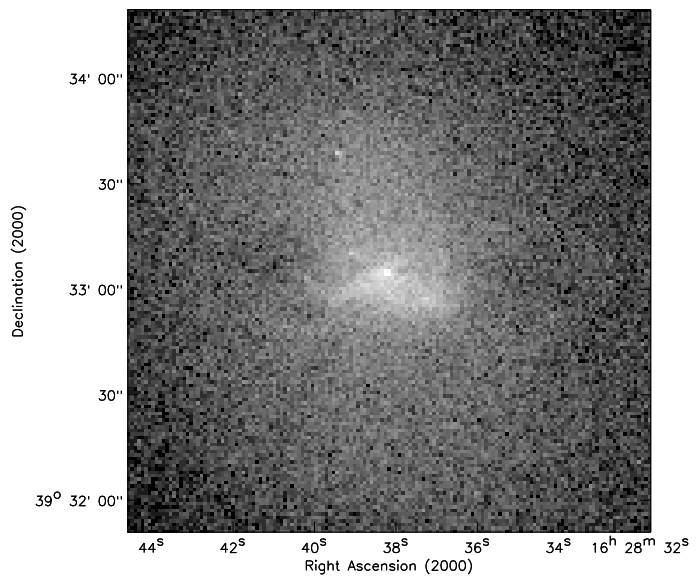}}
\caption{Raw X-ray image in the 0.3-7.0keV band. The data have been
  binned to 1 arcsec bins.}
\label{xray-raw}
\end{figure}

\begin{figure}
\protect\resizebox{\columnwidth}{!}
{\includegraphics{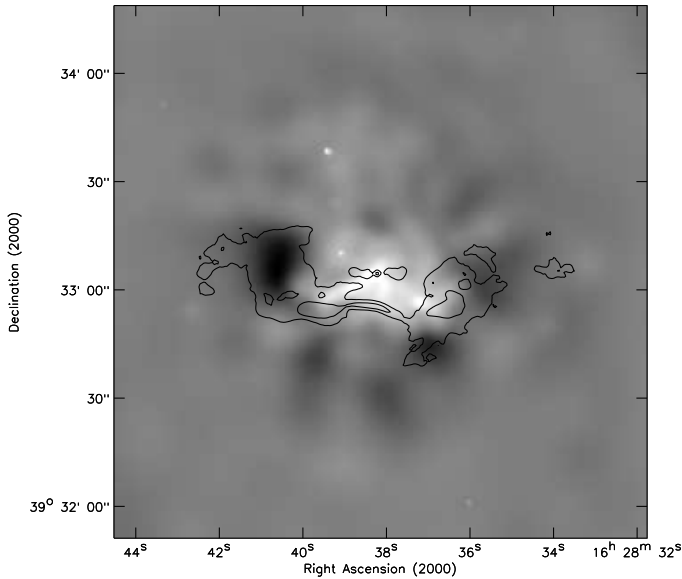}}
\caption{
Contours from a 1.7 GHz radio map \citep{Giovanninietal98} overlaid
on a difference image made from the X-ray data. The difference image
is constructed by subtracting an image smoothed with a gaussian of
standard deviation of 20 arcsec from the data which has been
adaptively smoothed using a three sigma threshold. Surface brightness
deviations can be seen at 20-40 arcsec from the nucleus to the east, south and
west. The eastern radio lobe coincides with the eastern depression in
surface brightness.}
\label{xray-radio}
\end{figure}

\section{Observations}
\label{observations}
Two observations of Abell 2199 were made using the Advanced CCD
Imaging Spectrometer (ACIS-S) on the {\it Chandra} satellite.  Details
are given in Table~\ref{obslog}.  The standard reprocessed level-two
events files have been used for analysis after screening for
background flares. Throughout, we adopt a redshift for Abell 2199 of
$z=$0.0309 and a cosmology with $H_0=50\kmpspMpc$ and q$_0$=0.5,
giving an angular diameter distance of 176 Mpc and a luminosity
distance of 187 Mpc. The linear scale is 850pc per arcsec.

\begin{table}
\begin{center}
\begin{tabular}{ccccccccccc} \\
\multicolumn{1}{c}{Date} &
\multicolumn{1}{c}{Obsid} &
\multicolumn{1}{c}{Seq num} &
\multicolumn{1}{c}{FP Temperature} &
\multicolumn{1}{c}{Duration} \\
\multicolumn{1}{c}{} &
\multicolumn{1}{c}{} &
\multicolumn{1}{c}{} &
\multicolumn{1}{c}{($^\circ$C)} &
\multicolumn{1}{c}{(seconds)} \\
\hline
2000 May 13 & 497 & 800005 & --120 & 17319 \\
1999 Dec 11 & 498 & 800006 & --110 & 15771 \\

\end{tabular}
\renewcommand{\baselinestretch}{1.0}
\newline
\caption{
Log of observations. The columns show the observation date,
observation identification number, observation sequence number, focal
plane temperature and exposure time after the observation had been cleaned
of background flare events.
}
\label{obslog}
\end{center}
\end{table}

In Fig.~\ref{xray-raw} we show the {\it Chandra} X-ray image of the
central $\sim$1 arcmin radius region of Abell~2199 in the 0.3-7.0~keV
band.  Events from both observations have been combined, taking into
account the different aspect solution of the two data sets. The X-ray
image was binned to have 1 arcsec pixels. A faint X-ray point source
coincident with the active nucleus of NGC~6166 is seen at
RA=16:28:38.24, Dec=39:33:04.3 (we use equinox 2000.0 coordinates
thoughout). This has been the subject of a previous paper
\citep{DM01}.

In Fig.~\ref{xray-radio} we show, as the image, the result of
adaptively smoothing (using a three-sigma threshold) the X-ray data
and then subtracting the gaussian smoothed (with standard deviation 20
arcsec) data. This emphasises deviations from the large-scale
structure. Overlaid on this image is the 1.7~GHz radio map of 3C338
\citep{Giovanninietal98}.  Depressions in X-ray surface brightness are
seen to the east, south and west of the galaxy nucleus. There is a
clear correspondence between the position of the eastern radio lobe
and a marked depression in the X-ray surface brightness. The south
eastern and north western ends of the western radio lobe are also
associated with depressions in X-ray surface brightness.

For the case of the eastern radio lobe it may be that the
radio-emitting plasma displaces the hot gas as is seen in the Perseus
(Fabian et al. 2000) and other clusters (e.g. McNamara et al.  2000).
The other depressions to the south may indicate the positions of older
radio lobes.

\begin{figure*}
\resizebox{\columnwidth}{!}
{\includegraphics{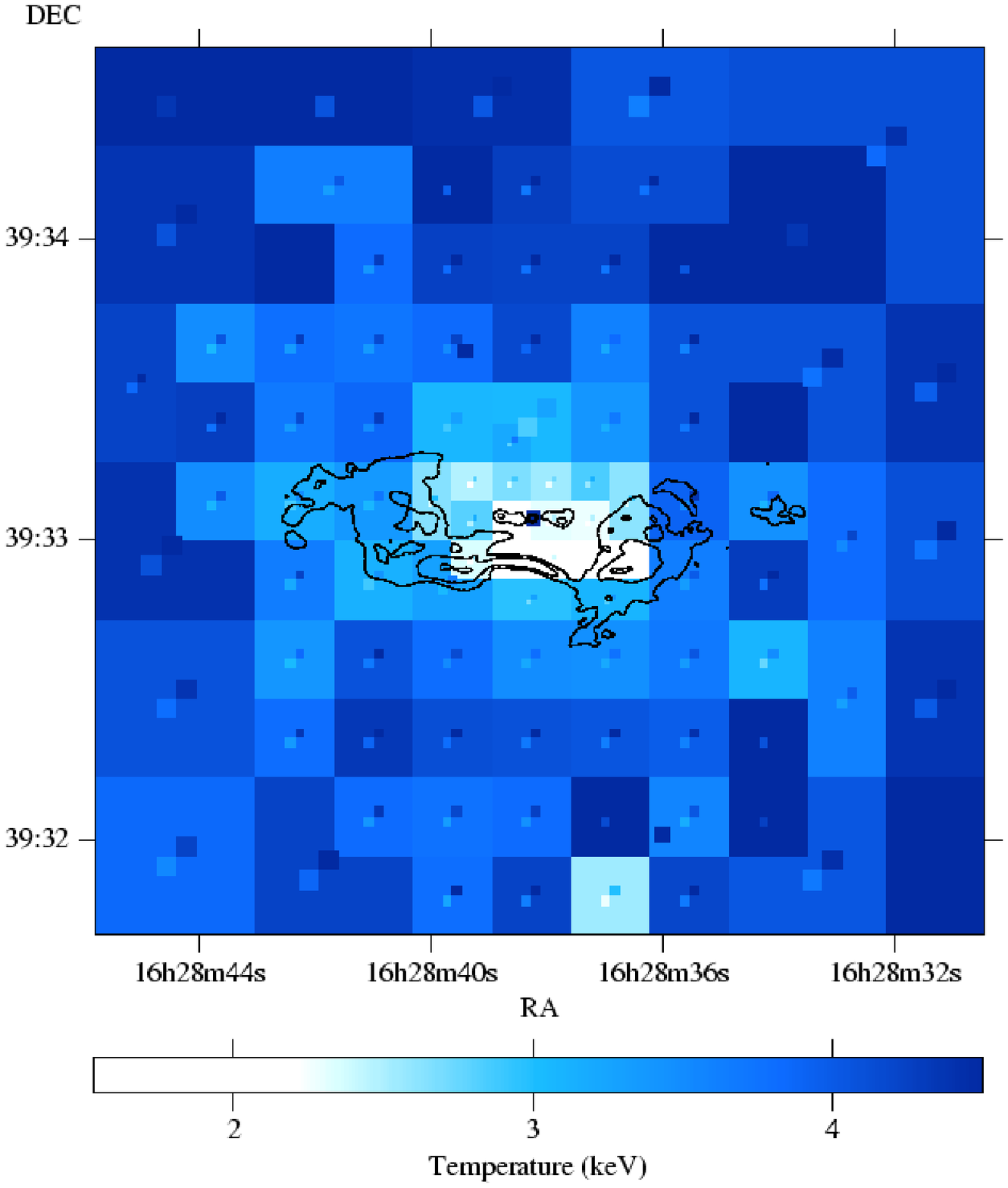}}
\hfill
\resizebox{\columnwidth}{!}
{\includegraphics{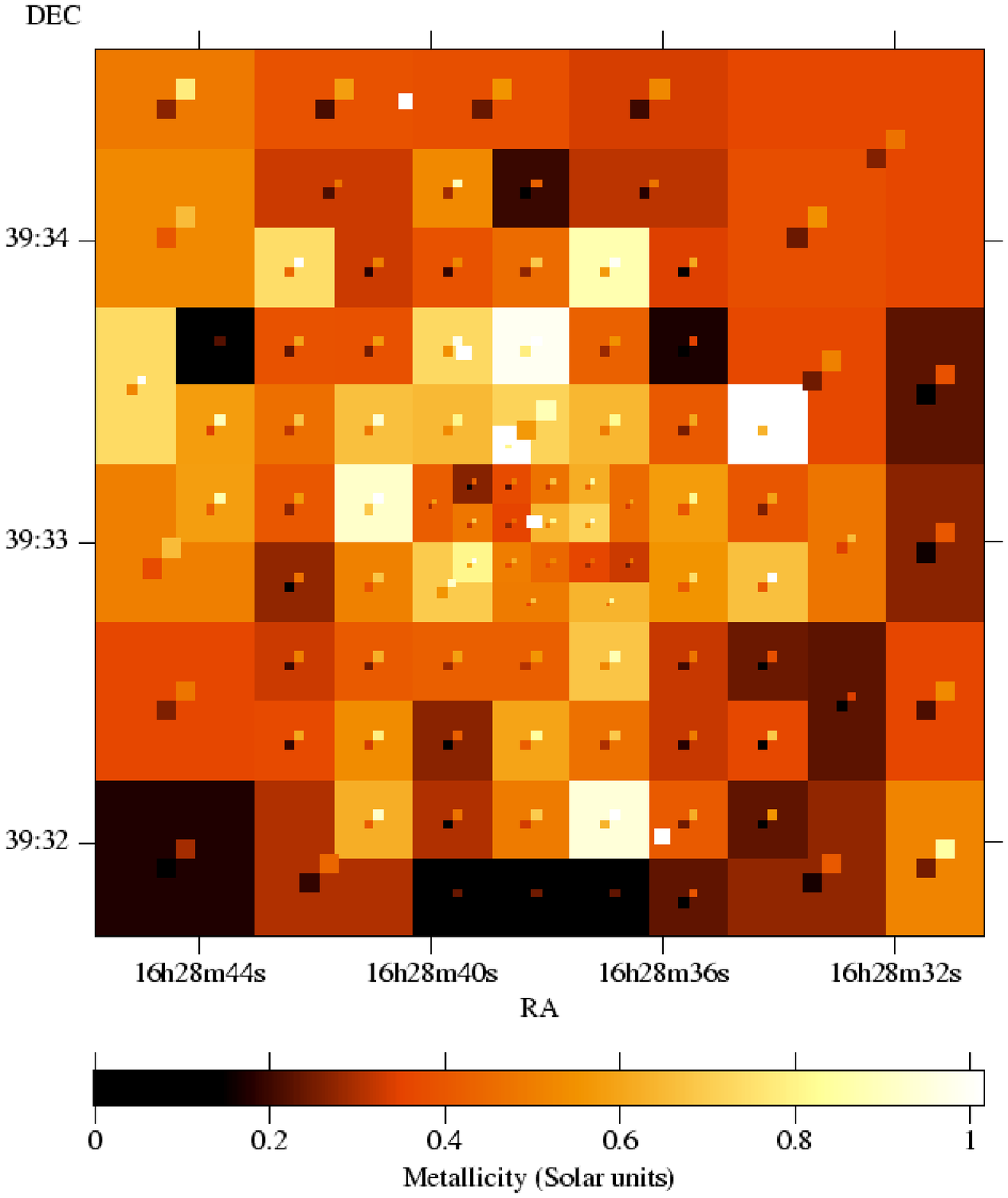}}

\resizebox{\columnwidth}{!}
{\includegraphics{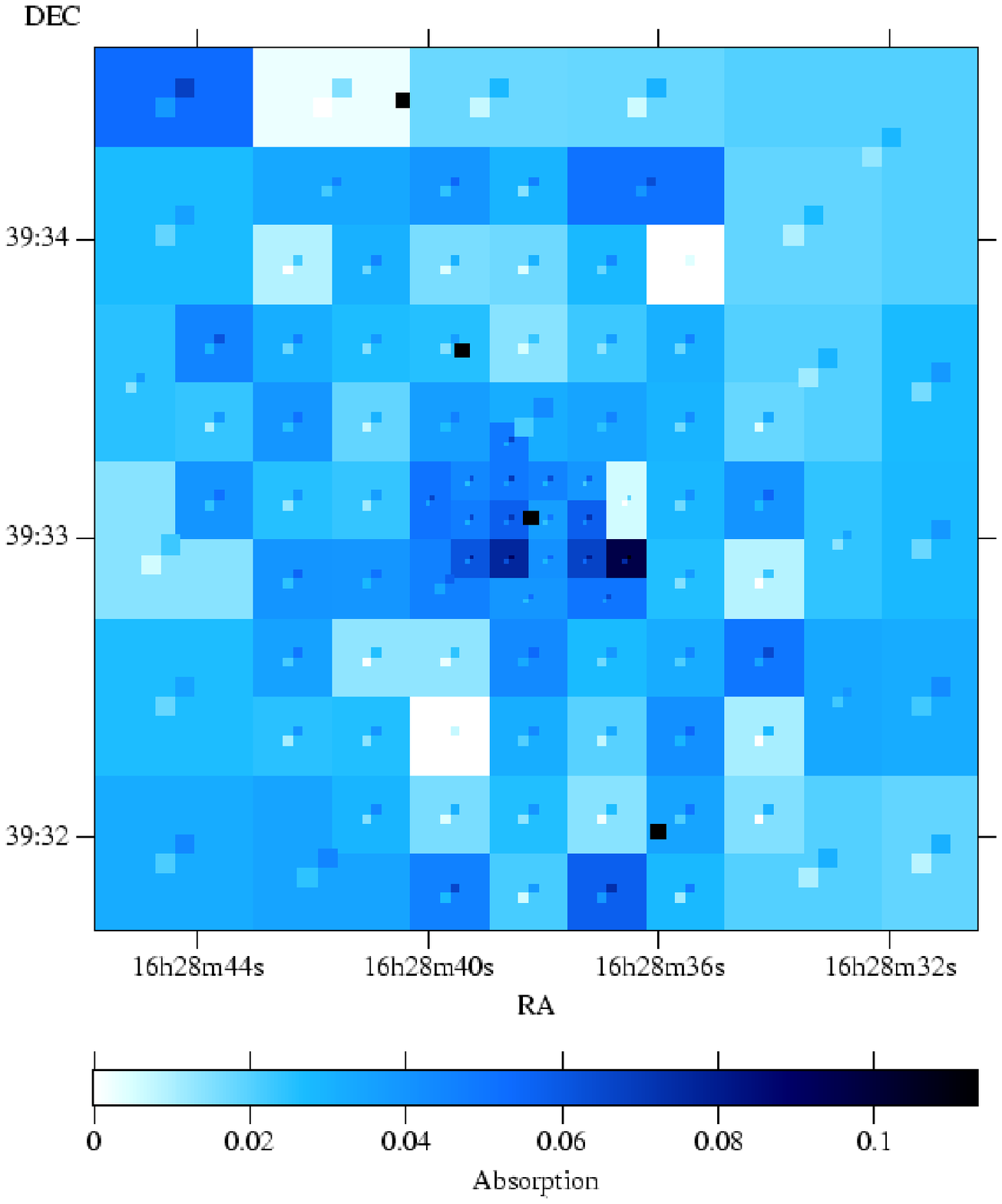}}
\hfill
\resizebox{\columnwidth}{!}
{\includegraphics{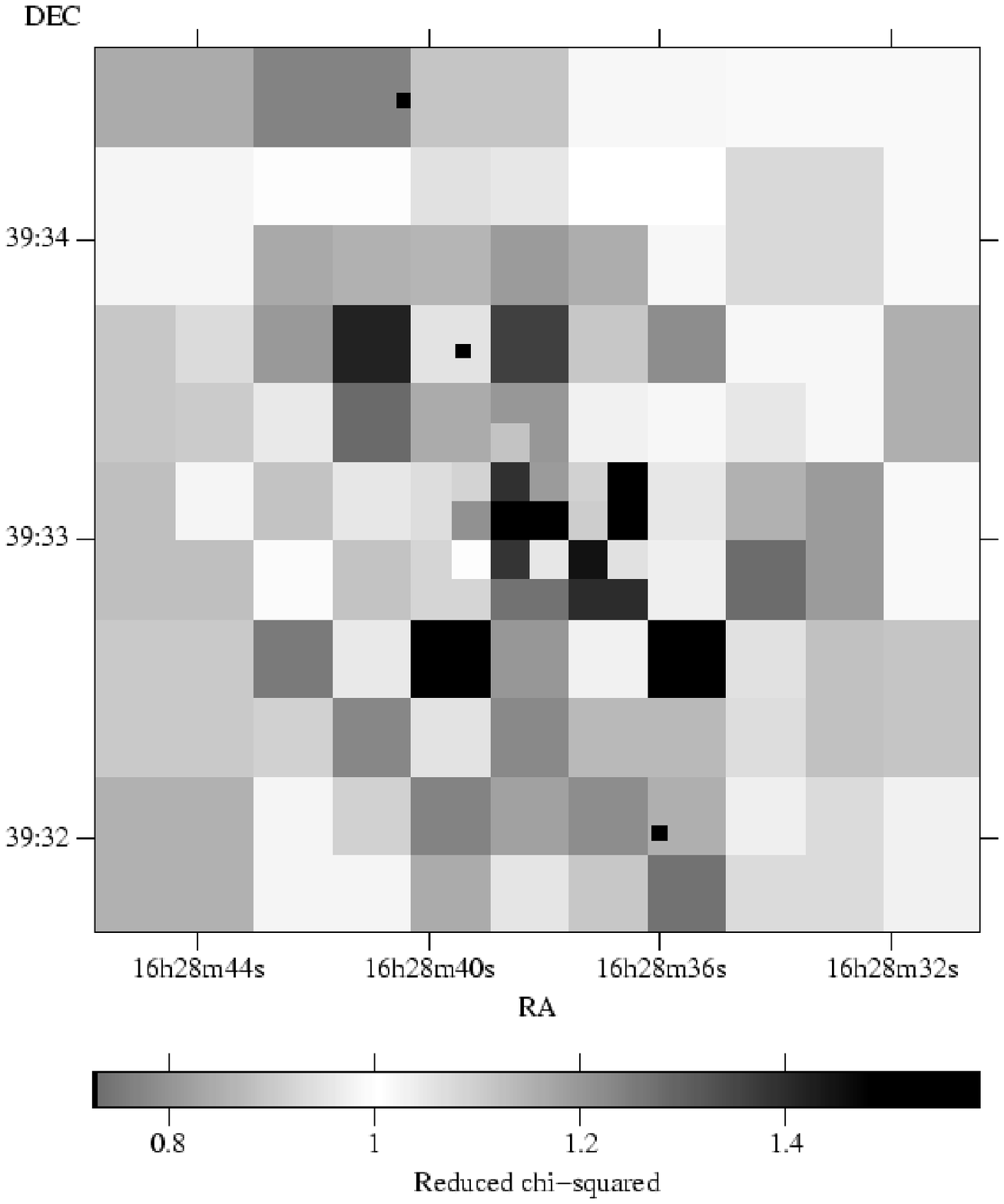}}
\caption{Temperature (with radio contours overlaid) (upper left), metallicity (upper right),
  absorption (in units of $\times 10^{22}\pscm$)(lower left) and reduced chi-square (lower right) maps of Abell
  2199 determined from spectral fitting of data extracted from
  adaptively binned
  spatial regions. The small dot pairs inside each bin show the $1\sigma$
  uncertainty on the values.
  Regions excluded from the analysis due to the
  presence of a point source are marked as small black or white squares.}
\label{jfigs}
\end{figure*}

The X-ray data do not account simply for the fact that the radio ridge
which joins the two radio lobes is displaced to the south of the
nucleus. It is plausible that motion of the nucleus is responsible for
this, as discussed by \citet{Burnsetal83}. Probably, it is not the
nucleus alone which moves but much of the whole central galaxy and
innermost cluster core. If, as seen in projection from Earth, the
galaxy has oscillated, or orbited, to the south a distance of about 10
to 20 arcsec (8.5--17~kpc) and has now returned to the north with the
radio source off for most of the return trip, then the radio
structure may be explained. The bright X-ray emitting region which
extends to the south of the nucleus could be a cooling wake, similar
to that seen in Abell 1795
\citep{Fabianetal01a}.

\section{Spatially resolved spectroscopy}
\label{spatialspectroscopy}

\subsection{Analysis of data in projection}
\label{projection}

The two observations of Abell 2199 were taken with the focal plane at
different temperatures, making it inappropriate to combine the event
data directly for spectral analysis. We have therefore produced
separate spectral files, response data and background files for the
two observations, and fitted models to the two observations
simultaneously using \textsc{xspec} \citep{Arnaud96}. 
Response matrices and effective area
files were made by averaging a $32\times32$ grid of calibrations
covering chip 7, using the \textsc{CIAO} tools \textsc{mkrmf} and \textsc{mkarf}, and using
weighting factors equal to the number of counts (in the 0.5-7.0 keV
band) in the source in the region covered by the calibration.
Background spectra were generated from corresponding regions of the
standard background fields using software and data from Maxim
Markevitch at
http://asc.harvard.edu/cal/Links/Acis/acis/Cal\_prods/\linebreak[0]bkgrnd/current/index.html.

Initially we did not wish to impose a particular geometry on the
possible spatial variation of spectral parameters. Therefore, we have
used a technique based on the adaptive binning algorithm of
\citet{Sanders01} to define tiled square regions of the
image where there are at
least 1000 counts by progressively increasing the binning of a spatial
map until that threshold of counts is reached. Pulse invariant spectra
were then extracted from these regions and fitted over the nominal
0.5-8.0 keV band with single temperature \textsc{mekal} \citep{Kaastra93,Liedahl95} plasma emission models that were allowed to have a
freely fitting value for the foreground absorption. (Discrete point
sources visible in the {\it Chandra} image at positions listed in Table \ref{pstable} have been excluded
from the analysis). We present the results of these fits in
Fig.~\ref{jfigs}.

\begin{table}
\begin{center}
\begin{tabular}{cc} \\
\multicolumn{1}{c}{RA} &
\multicolumn{1}{c}{Dec} \\
\multicolumn{2}{c}{(Equinox 2000)} \\

\hline
16:28:26.2 & 39:33:53 \\
16:28:38.3 & 39:33:04 \\
16:28:39.2 & 39:33:10 \\
16:28:39.5 & 39:33:38 \\
16:28:40.5 & 39:34:28 \\

\end{tabular}
\renewcommand{\baselinestretch}{1.0}
\newline
\caption{
Positions of point sources excluded from analysis.
}
\label{pstable}
\end{center}
\end{table}

The temperature map shows significant variations between adjacent
pixels. There is a trend in that the temperature drops from values of
4--4.5 keV around the outside edge of our map (R$\sim 75 \kpc$) to
values between 1.6--2.2 keV near the nuclear point source. The
significance of pixel-to-pixel variations in the metallicity map is
much less marked at the edge of our region (where values are between
0.15--0.3 of the solar value), due to the much larger percentage error
on this quantity. However, within the inner 45 kpc of the cluster
there are significant variations, with individual pixel metalicities
varying between 0.25 and 1.2 times the solar value. The absorption map
shows that the equivalent neutral hydrogen column density is at least
$2\times 10^{20}\pscm$ over most of the mapped region. Only a few
pixels are consistent with the value expected from our Galaxy for
which $N_H=8.7\pm0.6\times10^{19}\pscm$ (unweighted average of
surrounding grid points; \citealt{Dickey90}) in the direction of Abell
2199. Near the centre of the cluster the absorption rises to between
 $5-10\times 10^{20}\pscm$. Finally the map
of reduced chi-square shows that most of the pixels have values of
chi-square between 0.8 and 1.2. Only a few pixels, predominantly near
the centre of the cluster have higher values.

We have overlaid the radio contour map on the temperature map, to
facilitate comparison.  Although no detailed correlation is evident
between structure in the temperature or metallicity maps and the radio
map, the cooler regions of the cluster are associated with the radio
brighter parts of the radio source.

\begin{figure}
\protect\resizebox{\columnwidth}{!}
{\includegraphics[angle=-90]{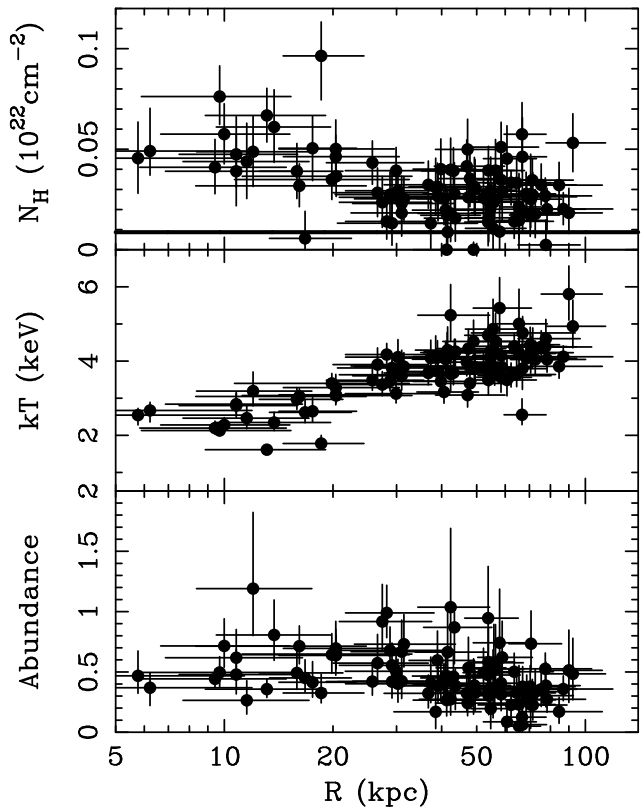}}
\caption{Radial variation of absorption, temperature, metallicity
in Abell 2199 as measured from a single temperature model fitted to the
adaptively sized bins. The solid line in the absorption
panel shows the Galactic value of the neutral hydrogen column density.
}
\label{jprof}
\end{figure}

In Fig~\ref{jprof} we have plotted the values of absorption column
density, temperature and metallicity as a function of the radial
distance of the bin from the cluster centre. Inspection of this plot
confirms that most bins at similar radii have similar properties
although there are a small number of values inconsistent with the mean
for that radius.  In particular, we have looked at the region near the
two cool pixels centred on (RA=16:28:36.9, Dec=39:32:55.8). We find
that a single temperature fit gives kT$\sim 1.3\keV$, rather cooler
then the surrounding regions.

\subsubsection{Single temperature radially varying models}
\label{1tmods}
In the absence of evidence for gross non-radial variations in the
properties of the gas we have made spectra binned azimuthally into eight
concentric rings out to a radial distance of 200 kpc from the central point source.
One further irregular shaped bin is included which contains all of the 
data that are outside 200 kpc radius region but which still lie
on chip 7. Bins beyond a radial distance of 140 kpc from the
nucleus cover slightly different spatial regions in the two data sets,
because the cluster is not centred on the chip and the 
spacecraft roll angle was different for the two observations.

\begin{figure}
\protect\resizebox{\columnwidth}{!}
{\includegraphics{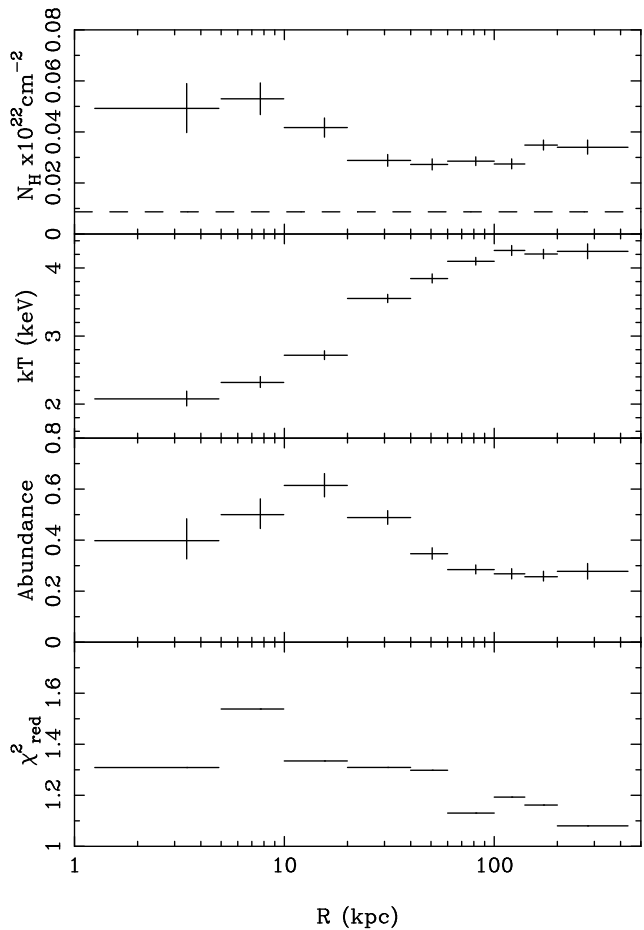}}
\caption{Radial variation of absorption, temperature, metallicity and
reduced chi-square in Abell~2199. The dashed line in the absorption
panel shows the Galactic value of the neutral hydrogen column density.
}
\label{rmjprof}
\end{figure}

We have fitted a single temperature \textsc{mekal} model absorbed by a
foreground screen (to account for absorption from the Galaxy) to each
of the concentric spatial regions.
Fig.~\ref{rmjprof} shows the radial profiles of
temperature, metallicity, equivalent hydrogen absorbing column density
and reduced chi-square.
The absorption, parameterised as the equivalent column density of
neutral hydrogen, is greater than that expected from 21cm observations of the
Galaxy thoughout the region studied. (This may be related to
uncertainties in the
calibration of the instrumental effective area below 1keV\footnote{
http://cxc.harvard.edu/cal/Links/Acis/acis/Cal\_projects/index.html
}).
Even with this caveat, though, we see an increase in the
absorption within the central 20 kpc, relative to that seen further out,
by $\sim2\times10^{20}\pscm$.

The cluster shows a smoothly declining temperature profile that starts
at $\sim 4.2$ keV beyond 200 kpc, reducing to near 2 keV within the central 5
kpc. The broad-beam temperature for this cluster was measured with
EXOSAT as $4.7\pm0.4$ keV by \citet{EdgeStewart91} indicating that the
temperature profile climbs further beyond the spatial extent of our
analysis. Our outer measured temperature is in good agreement with the value
of $4.16\pm0.15$keV measured from the central 4-5 arcmin radius region by ASCA
\citep{Allen01a}.

Within the central 5kpc we find that the metallicity is close to the
ASCA measured value for the central 4-5 arcmin radius region \citep{Allen01a} of
0.4 times the solar value. (We use the default solar abundance set given
by \citealt{Anders89}). By a radial distance of 15kpc, the metallicity has risen to
0.6 times the solar value, before reducing again at larger radii to near 0.25
times the solar value.

\subsubsection{Multi-temperature models}
\label{2tmods}
We have also examined the annular spectra using multi-temperature models. 
We first fitted the data with a two-temperature model consisting of
freely fitting absorption acting on two single-temperature emission components;
the abundances were tied
together but the temperatures were allowed to fit freely.
The introduction of the second emission component is required (at the
99 per cent confidence level, determined using an F-test)
for the three radial bins 10--20kpc,
20--40kpc and 140--200kpc.
However, in section \ref{spectraldeprojection} we show that the evidence for
multi-temperature components in these annular regions
is largely due to projection effects.

Recently, \citet{protassov2002} have questioned the validity of
using the F-test to test for the significance of an additional
additive component when fitting models to data because the hypothesis
is on the boundary of the parameter space. In our case we have used
the F-test to check whether a second \textsc{mekal} component is statistically
required by the data.  This is on the boundary of parameter space
since the null hypothesis is that the second \textsc{mekal} component is not
required. In this case the normalization of the second component is
zero, and negative normalizations (although possible in practice
within the fitting code) are excluded from the fitting procedure as
they are not physically meaningful. We have carried out fits to
Monte Carlo realizations of our data and find that (in our case) the
distribution function of our measured F values matches the formal F
distribution very well for values of F above 0.2\footnote{

To check the validity of the F-test we have repeatedly run fits with
single and two-temperature \textsc{mekal} models to Monte Carlo realizations of
our data (generated using the \textsc{xspec fakeit} command) and looked at the
distribution of the F-statistic. One thousand realizations of our data
were generated from the single temperature plasma model (with freely
fit absorption) which best fitted one combined data set, by folding the
model through the detector response function and adding random noise
appropriate to the total counts generated. First we fitted the
same model that was used to generate the data and noted the chi-square
value and number of degrees of freedom. We then fitted a
two-temperature \textsc{mekal} plasma model in which the metallicity of the
second component was fixed to that of the first and the absorption
component acted on both \textsc{mekal} components. The temperature and normalization
of the second component are then the new fit parameters; we logged the value of
the chi-square statistic and number of degrees of freedom in this case.

A Kolmogorov-Smirnov test was used to determine whether the
distribution of measured F values was significantly different from the
true F distribution. For all values of F above 0.2 we find the two
distributions to be in excellent agreement (the probability of
differences between the two cumulative distribution functions as large
as observed (if they are the same distribution) is greater than 20 per
cent).

Below a measured F-value of 0.2 the two distributions are significantly
different (much less than 1 per cent chance of deviations as large as
observed if the two distributions are the same). This may be due to
either to \textsc{xspec} sometimes not finding the exact minimum value of
chi-square, or to problems with the test discussed by \citet{protassov2002}.

Since the critical values of F used in our assessment of the
significance of extra spectral components in this paper lie near
F=4.6, we conclude that the boundary condition problem does not affect
our inferences.

}.

We next explored the effect of replacing the second temperature
component with a cooling flow spectrum (in which gas cools
to zero degrees
from the plasma temperature). We obtain similar results to
those found with the two-temperature model and place limits on the
size of the cooling flow component within R=100kpc of $12\pm3\Msunpyr$
or $37^{+14}_{-13}\Msunpyr$ (90 per cent confidence limits), 
depending whether or not freely fitting intrinsic absorption is
included on the cooling flow component.

We have also investigated the region, mostly to the south
of the  nucleus, defined by the polygon with the following vertices:
(16:28:40.03,39:32:54.7),\linebreak[0]
(16:28:40.03,39:32:54.7),\linebreak[0]
(16:28:38.17,39:33:12.9),\linebreak[0]
(16:28:36.25,39:32:54.2),\linebreak[0]
(16:28:40.05,39:32:54.1), but excluding a region of
radius 2.5 arcsec centred on the nucleus. This region was targeted for
separate analysis because it stands out as having a higher
surface brightness than surrounding regions. We find that the gas
temperature (fitted with a single-temperature model) is cool
(T$=2.5\pm0.1$ keV) and that the data can accommodate a cooling flow
of up to $9\Msunpyr$.
Since this triangular region contains regions of bright radio emission
we have checked for the presence of power-law continuum X-ray emission.
No significant such emission was found: a two temperature fit is
significantly better than a single temperature plus power-law continuum.

\subsection{Analysis of data accounting for projection effects}
\label{deprojection}
\subsubsection{Single temperature fits}
\label{spectraldeprojection}

Assuming that the cluster is spherically symmetric and that the
physical conditions change only in the radial coordinate of the
cluster, we can account for the projection effects of gas exterior to
a particular region. This allows us to determine the physical
conditions in the gas as a function of the true 3-dimensional radial
coordinate.  We do this by fitting spectra from concentric rings
simultaneously using multiple \textsc{mekal} models.  We allow one
\textsc{mekal} model for the outermost annulus, two for the next one
in and so on, setting the normalizations of the different projected
components according to the relative volumes of the shells. The
technique is described in full in \citet{Allen01b}. Another approach,
used on the cluster Abell 1795, is described by \citet{Ettori02}.

\begin{figure}
\protect\resizebox{\columnwidth}{!}
{\includegraphics{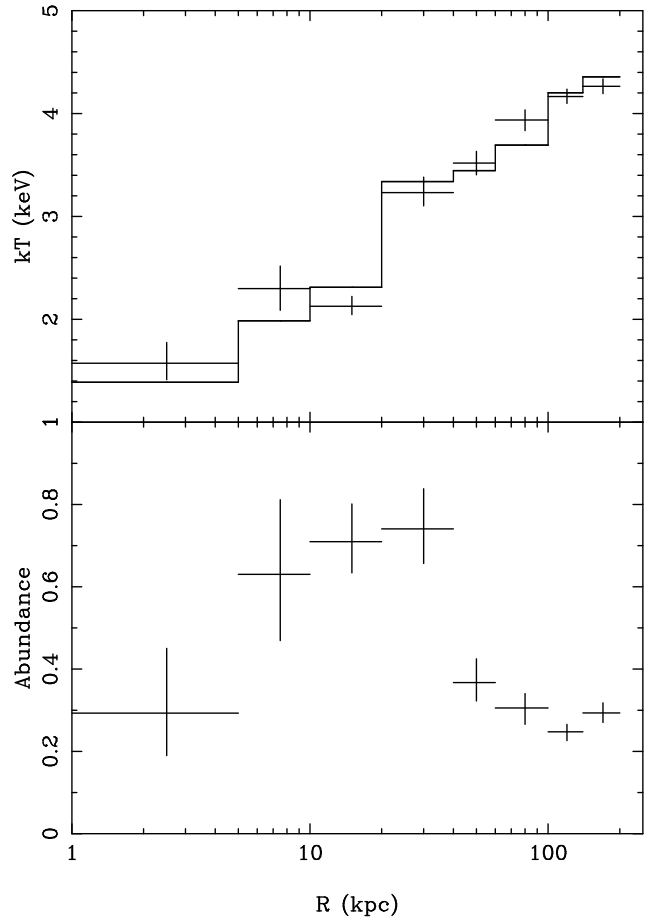}}
\caption{Intrinsic (3-dimensional) radial variation of temperature and
metallicity in Abell 2199. The solid line plotted over the temperature
profile is the \citet{Navarro97} (NFW) profile which best fits the data.
}
\label{rmjdeprojdataprof}
\end{figure}

The temperature and abundance profiles obtained from this method are
shown in Fig.~\ref{rmjdeprojdataprof}. The Galactic absorption was allowed to
be a free parameter and its value was determined as $3.05\pm0.09\times10^{20}
\pscm$. The temperature profile shows the central bin to be
cooler than that measured in the projected data: $1.57^{+0.20}_{-0.16}$ keV
compared with $2.08^{+0.12}_{-0.10}$ keV. The
profiles both from the simple analysis presented in section \ref{1tmods}
and from this analysis asymptote to similar values, $4.26\pm0.07$keV
(deprojected) and $4.21\pm0.07$keV (seen in projection) at radii of 140-200kpc.

The deprojected abundance profile is very similar to that seen in the
simple single-temperature fits to the data in projection.
The low value seen in the central bin, and the value seen at the
abundance maximum near 20 kpc, are more extreme in the deprojected
analysis but with larger uncertainties.

\subsubsection{Two temperature fits}
\label{spectraldeprojection2t}

Finally, we have investigated whether there is a statistical requirement for a
second emission component in any annulus, once the projection of
the first emission components have been taken into account. To do this,
we use the \textsc{xspec projct} model and add in second emission components to each
annulus successively (starting at the outside, allowing only the
temperature and normalizations to be freely fit), and
note how the value of the chi-square statistic
reduces. In each case the metal abundance of the second component is tied to
the metal abundance of the corresponding first component, and a single
absorption component is applied to the whole model.
The fiducial fit, in which each region has only one emission
component, gave a value $\chi^2=4865.3$ for 3761 degrees of freedom.
Table \ref{2tdeltas} shows the values of the chi-square statistic, number of
degrees of freedom and corresponding values of
the F-statistic, together with the formal probability that such a
large value of F 
would be obtained if the new emission component was not
required. We caution that due to the large value of the reduced
chi-square statistic the true probabilities will be larger than listed
in the table, so the results are less significant than tabulated.

\begin{table}
\begin{center}
\begin{tabular}{ccccc} \\
\multicolumn{1}{c}{Region} &
\multicolumn{1}{c}{$\chi^2$} &
\multicolumn{1}{c}{DoF} &
\multicolumn{1}{c}{F} &
\multicolumn{1}{c}{F$_{prob}$} \\
\multicolumn{1}{c}{(kpc)} \\

\hline
140-200 & 4833.0 & 3759 & 12.6 & 3.7E-6 \\
100-140 & 4823.4 & 3757 & 3.7  & 2.4E-2 \\
60-100  & 4802.7 & 3755 & 8.1  & 3.1E-4 \\
40-60   & 4788.5 & 3753 & 5.6  & 3.9E-3 \\
20-40   & 4770.5 & 3751 & 7.1  & 8.6E-4 \\
10-20   & 4761.2 & 3749 & 3.6  & 2.6E-2 \\
5-10    & 4755.3 & 3747 & 2.3  & 9.6E-2 \\
0.8-5   & 4750.9 & 3745 & 1.7  & 1.8E-1 \\

\end{tabular}
\renewcommand{\baselinestretch}{1.0}
\newline
\caption{
Values of $\chi^2$, number of degrees of freedom, the F statistic and
corresponding probability for including a second emission component in
each region successively, starting from the outermost region.
}
\label{2tdeltas}
\end{center}
\end{table}

We find that the outermost region strongly requires a second emission
component. This is not surprising since by definition we have not
accounted for the projection of emission external on to this region.
There is also evidence, but at a lower significance, for a second
emission component being required between 20-100kpc.

\section{Image deprojection and Mass Analysis}
\label{imagedeprojection}

The deprojected temperature and surface brightness profiles can be
used to 
constrain the mass profile of the cluster. For this analysis we have used 
the methods described by \citet{Allen01b} and \citet{schmidt01}.
In brief, the observed surface brightness  profile and a
particular parameterized mass model are together used to  predict the
temperature profile of the X-ray gas, which is then compared with 
the observed, spectrally-determined results. 

Three separate parameterizations 
for the cluster mass distribution were examined: the 
\citet{Navarro97} (NFW) profile, the Moore et al. (1998) profile and a
non-singular isothermal sphere (NIS). We find that the NFW
profile provides (marginally) the best fit to the data, although all
three models give formally unacceptable results. 
Table~\ref{massres} lists the best-fit parameters
for the three models, together with their $1\sigma$ uncertainties, 
chi-square values, and the number of degrees of freedom for the fit. 
For the NFW and Moore profiles we note that there are only two fit
parameters; the scale radius, $r_{\rm s}$, and the concentration parameter, 
$c$. However, we also quote values for the effective velocity dispersion
$\sigma= \sqrt{50} r_{\rm s} c H$, which is a function of the other two 
variables and the Hubble constant. 

\begin{figure}
\rotatebox{270} {
\resizebox{!}{\columnwidth}
{\includegraphics{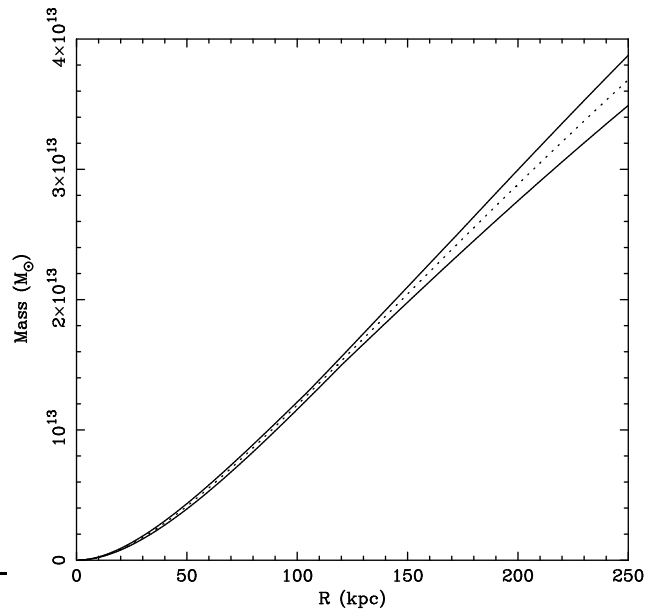}}}
\caption{Mass profile (dotted line) of the core of Abell 2199 determined from the NFW
potential which best fits the temperature profile. The solid lines show the
$1\sigma$ uncertainty limits.
}
\label{massprof}
\end{figure}

The temperature profile predicted by the best-fit NFW mass model
overplotted on the observed, deprojected temperature profile is shown 
in the top 
panel of Fig.~\ref{rmjdeprojdataprof}. The total gravitating mass profile 
associated with the best-fit NFW model is
plotted in Fig.~\ref{massprof}, along with
its $1\sigma$ uncertainties.

\begin{table}
\begin{center}
\begin{tabular}{ccccccccccc} \\
\multicolumn{1}{c}{NFW} &
\multicolumn{1}{c}{Moore} &
\multicolumn{1}{c}{NIS} \\

\hline
r$_{\rm s}=130\pm10\kpc$&r$_{\rm s}=650^{+250}_{-150}\kpc$&r$_{\rm c}=20^{+1}_{-2}\kpc$\\

c$=9.7\pm0.5$&c=$2.6^{+0.5}_{-0.6}$ \\

$\sigma=467\pm12\kmps$& $\sigma=625^{+75}_{-70}\kmps$ & $\sigma=600\pm10\kmps$\\

$\chi^2=15.9/6$DoF & $\chi^2=23.7/6$DoF &$\chi^2=20.2/6$DoF\\

\end{tabular}
\renewcommand{\baselinestretch}{1.0}
\newline
\caption{
Parameters obtained from fitting mass models to the temperature profile.
}
\label{massres}
\end{center}
\end{table}

The shape of the inner gravitational well around NGC6166 has been probed
by the optical spectroscopy of \citet{Carteretal99}, who find
that the velocity dispersion of the galaxy rises outward. Recently
\citet{Kelsonetal02} have confirmed and extended this result. They
found that the velocity dispersion continues to rise out to the limit
of their data at 60kpc, by which time the stellar velocity dispersion is close
to that of the cluster galaxies. \citet{Kelsonetal02} note that
our measurement of the mass within $0.1h^{-1}$Mpc (where $h=H_0/100$) is
in reasonable agreement with their determination, for the same region.

We next used the best-fitting gravitational potential that was
determined above to constrain an image
deprojection analysis of the Abell 2199 cluster. We follow the method of
\citet{Fabian81} in determining the physical conditions in the cluster as
a function of the radius. The Galactic value of the column density is used.
Error bars for the derived quantities were determined from one hundred
Monte Carlo replications of the original data.

\begin{figure}
\resizebox{\columnwidth}{!}
{\includegraphics[angle=-90]{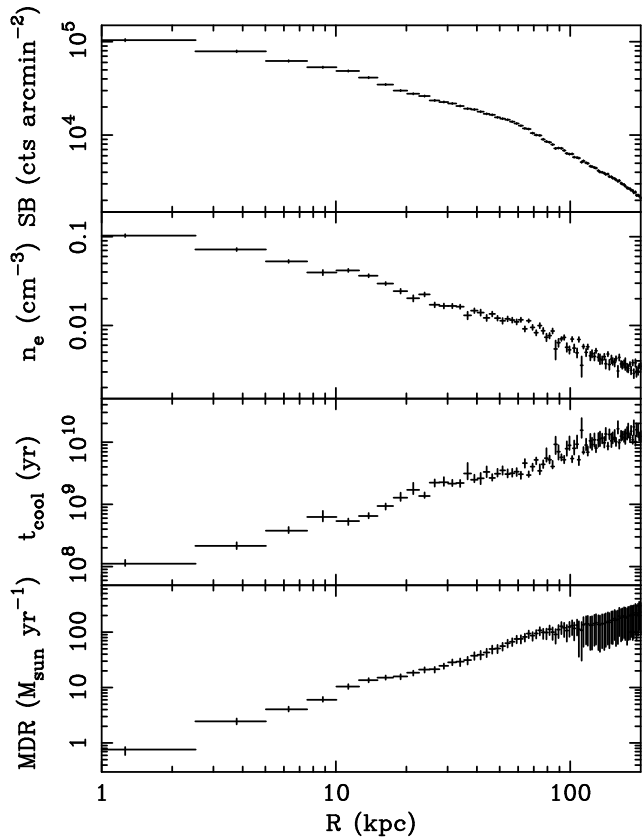}}
\caption{Image deprojection results of Abell 2199. The top panel shows the input surface
brightness profile, while the lower three panels show the derived electron density, cooling
time and mass deposition rate profiles in the cooling flow.
}
\label{sprojprof}
\end{figure}

In Fig.~\ref{sprojprof} we show the major results from the
deprojection analysis. The input surface brightness distribution is
shown in the top panel together with the implied electron density,
cooling time and mass deposition rate in lower panels.

The nominal mass deposition rate integrated out to the point where the cooling
time is equal to the age of the universe is $150
\Msunpyr$. However, the spectral fits (section 3.1.2) place a limit on the mass
deposition rate of less than $\sim50\Msunpyr$ suggesting that any steady
state cooling flow must be younger than 2--3~Gyr (associating the age
of the cooling flow with the cooling time of
gas where the mass deposition rate from the image deprojection equals
that in the spectral fits). 

\section{Heating and cooling rates}

We have seen that the radiative cooling time of the gas in Abell 2199 is
less than about 7~Gyr within a radius of 100~kpc, less than
2~Gyr within 20~kpc and about 0.1~Gyr within the innermost few kpc.
The temperature decreases from above 4~keV beyond 100~kpc to 1.6~keV
(deprojected) within 5~kpc. This overall appearance suggests that a
cooling flow may be operating (see e.g. Fabian 1994), but the spectral
analysis provides only modest support and then only for a small flow
of tens $\Msunpyr$. XMM-Newton Reflection Grating Spectrometer (RGS)
data of other clusters also show
little evidence for long-lived, steady-state
cooling flows, in which gas cools from
the cluster virial temperature to below $10^6\K$ by emitting observed
X-radiation (Peterson et al. 2001; Tamura et al. 2001; Kaastra et al.
2001). The RGS spectra, which do not spatially resolve any flow, are
consistent with the gas cooling from the virial temperature to about
one third to one fifth of that value but no lower, possibly in a
single-phase cooling flow. The fate of the coolest gas, at 1--2~keV,
is unknown. It may be heated, mixed with much cooler gas or expelled
(see e.g. Fabian et al. 2001b).

\begin{figure*}
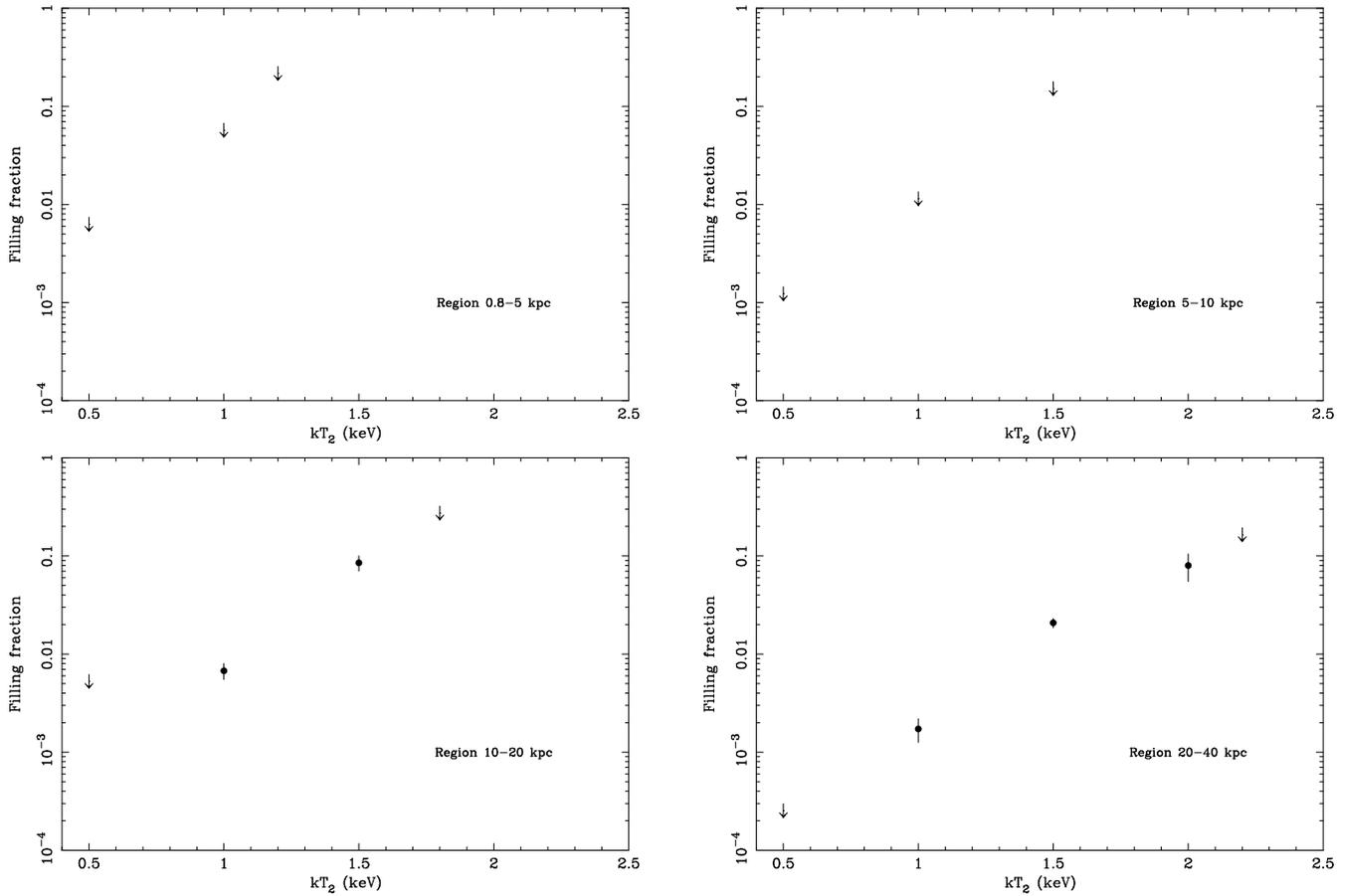

\rotatebox{270}{
\resizebox{!}{\columnwidth}
{\includegraphics{f_ring0.ps}}}
\hfill
\rotatebox{270}{
\resizebox{!}{\columnwidth}
{\includegraphics{f_ring1.ps}}}

\rotatebox{270} {
\resizebox{!}{\columnwidth}
{\includegraphics{f_ring2.ps}}}
\hfill
\rotatebox{270} {
\resizebox{!}{\columnwidth}
{\includegraphics{f_ring3.ps}}}
\caption{Volume filling fractions of a second temperature component,
  in the innermost four spatial
regions, as a function of temperature.}
\label{fillfact}
\end{figure*}

\begin{figure}
\rotatebox{270}{
\resizebox{!}{\columnwidth}
{\includegraphics{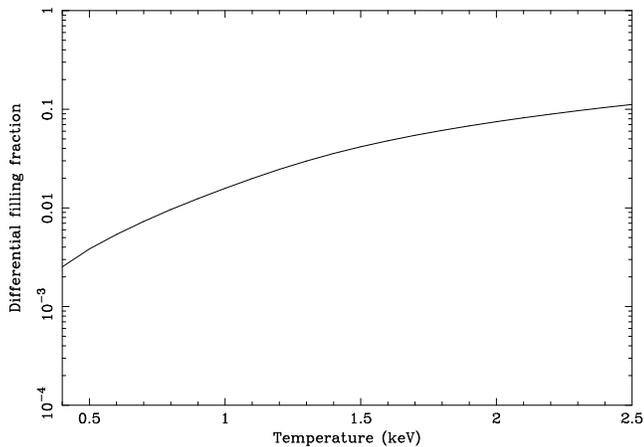}}}
\caption{Differential filling fraction distribution of gas cooling
  from 2.5keV at constant pressure}
\label{cpcffillfact}
\end{figure}

Here we attempt to quantify upper limits to the amount of multiphase gas
in the core of Abell 2199, and to see where the steady-state cooling flow
assumption breaks down. Using a similar
(\textsc{projct}) spectral model, accounting for the projection of outer components
on to inner regions, to that discussed in section
\ref{spectraldeprojection2t},
limits on the volume filling fraction ($f_2$) of any second emission component
in the central four spatial bins (out to 40~kpc from the centre
of the cluster) were calculated using
$$f_2=\left[1+{e_1\over{e_2}}\left({T_1\over
      T_2}\right)^2\right]^{-1},\eqno(1)$$
\citet{Sanders02}. $e_1$ and
$T_1$ are the emission measure and temperature of the first
temperature component for a particular shell, and $e_2$ and $T_2$ are the
emission measure and temperature of the additional component. Unlike
section \ref{spectraldeprojection2t} we only applied the second
emission component to one spatial region at a time. The second
component has a freely fitting normalization and a metal abundance
that is fixed to the metal abundance of the first component. We
stepped the temperature of the second component through a range of
values from 0.5keV to 2.5keV, calculating $f$ and its uncertainty $\sigma_f$
(obtained by propagating the uncertainties in $e1$, $e2$, and $T_1$). 
Where the value of $f$ is less than $3\sigma_f$ we plot an upper limit
at $3\sigma_f$. Our
results are shown in Fig.~\ref{fillfact}. Also shown, in
Fig.~\ref{cpcffillfact}, for comparison, is the continuous
differential filling
factor distribution expected from a constant pressure cooling flow.

It is clear that there is little evidence for the gas being multiphase
at any radius. (Note that the `detections' of a second component at
some temperatures in the 10-20kpc region may be in large part due to
the projection of the second component present in the 20-40kpc
region). For a continuous distribution of temperatures the y-axis is
$df/dT$ (Fig.~\ref{cpcffillfact}) and limiting curves lie below the
plotted temperature limits which represent the step by step values
for one additional temperature component only. (Note also that the
test applied in section \ref{spectraldeprojection2t} is different from
the test applied here, as in the former each additional component
added to outer rings was left in when evaluating the F-statistic for
the interior ring, whereas here we have added the second component in
for each region separately and then removed it before adding in the
second component for the next region).

\begin{figure}
\rotatebox{270} {
\resizebox{!}{\columnwidth}
{\includegraphics{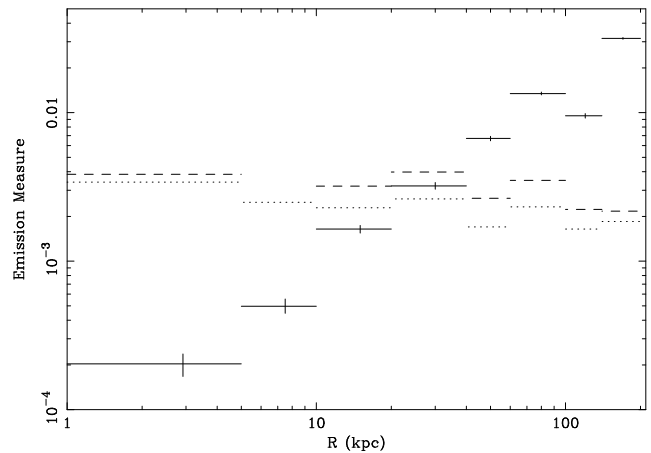}}}
\caption{The emission measure (in \textsc{xspec} units) of gas in each annulus,
determined from the deprojected spectral fitting procedure (solid
points). The dashed
curve shows the emission measure distribution expected from
single-phase gas cooling in the inferred gravitational potential at a
rate of $100\Msunpyr$. For comparison the dotted curve shows the
result for an isobaric, single-phase flow.
}
\label{em}
\end{figure}

Since the temperature distribution appears to be close to single
phase, we have then tested whether it resembles a single-phase cooling
flow dropping to some minimum temperature $T_{\rm min}\sim 1\keV$.
Although the temperature profile (Fig.~\ref{rmjdeprojdataprof})
appears consistent with a single-phase cooling flow, the surface
brightness profile is not. This is the reason that multiphase flows
were introduced in the first place (e.g. Fabian, Nulsen \& Canizares
1984). Fig.~\ref{em} (solid points) shows the distribution of emission measure
against radius. We use the normalizations from \textsc {xspec}, which need to be
multiplied by $3.8\times 10^{68}$cm$^2$ to convert to $\pccm$.

Also plotted is the distribution expected from a single phase cooling
flow of $100\Msunpyr$, both with and without the gravitational work
done included. To obtain this we use the energy equation of a steady
state spherical inflow,
$$-\dot M\int\left({5\over2}{k\over{\mu m}}{{d T}\over{d r}}+{{d
\phi}\over{d r}}\right) dr = -e\Lambda +h.\eqno(2)$$
$\mu m$ is the mean mass per particle, $h$ the heating rate (assumed
zero for Fig.~\ref{em}), $e$ the emission measure and $\phi$ the
gravitational potential, obtained from Fig.~\ref{massprof}. 

We see that the observed emission measure distribution is nowhere as
flat as expected from a single-phase flow (the expected values scale
up and down with $\dot M$).  There is too little emission at small
radii and too much at large radii. The latter can be explained by any
flow not yet having reached a steady-state, but the lack of emission
at the centre is more difficult to understand.  We have already shown
that spectral fits to the data exclude any simple distributed mass
dropout such as expected from a multiphase flow with a mass deposition
rate greater than $15\Msunpyr$.  However, a flow can be accommodated
if the gas, or its emission, can be made to be unobservable. Including
absorption is one way to achieve this, and the spectral data allow up
to $\sim 50\Msunpyr$ if the emission is absorbed. The missing soft
X-ray luminosity in cooling flows is discussed more extensively by
\citet{FabmissXray}.

\begin{figure}
\rotatebox{270} {
\resizebox{!}{\columnwidth}
{\includegraphics{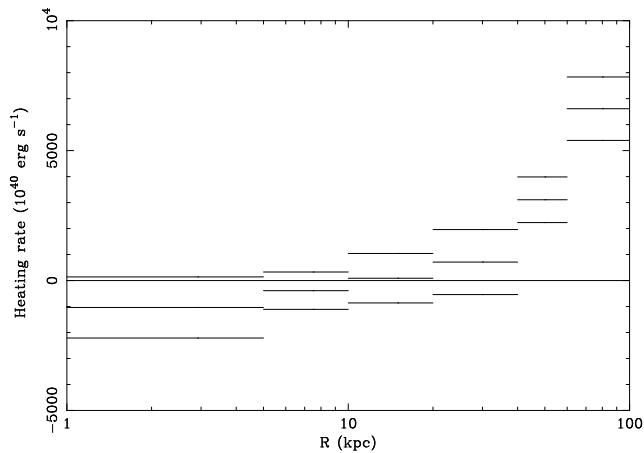}}}
\caption{The heating rate required to maintain thermal equilibrium for
a single phase inflow of 0 (upper lines), 50 (middle lines) and
$100\Msunpyr$ (lower lines).
}
\label{heat}
\end{figure}

Finally, we have quantified the heating required to maintain the gas
in a steady state (Fig.~\ref{heat}), by estimating $h$ from equation
(2). Here a negative value for the heating means that heat has been
lost in a non-radiative manner (for example by mixing with cold gas).
{\it Heating must be distributed over a wide range of radii}, with a
heating rate per unit volume proportional to $r^{1.5}$ if there is no
mass dropout. The results are only relevant where some steady-state
can be assumed. Note that the radio source, at present, affects the gas
distribution in the east-west axis out to about 30~kpc.

The peaked abundance distribution determined from the deprojected
spectral analysis, Fig.~\ref{rmjdeprojdataprof}, argues against much
mixing, or convection of gas across the annulus at 20--30~kpc. The
whole of the current gas distribution is stable against convection
since the entropy, $S\propto T/n^{2/3}$, monotonically increases with
radius ($S\propto r^{0.8}$ from a few kpc to about 100~kpc since the
deprojected gas temperature is approximately $1.11r^{0.29}\keV$ and
the electron density is $0.21 r^{-0.75}\pccm,$ where the radius $r$ is
in kpc).

\section{Conclusion}

The core of Abell 2199 is complex. Some depressions in the X-ray
surface brightness are seen to coincide with the radio lobes. Other
depressions to the south may indicate earlier radio activity. The
temperature decreases from 4.2~keV to 1.6~keV over radii from 100~kpc
to 5~kpc, which is where the radiative cooling time drops from 7~Gyr
to 0.1~Gyr. The X-ray surface brightness is not sufficiently peaked
however to be consistent with a single phase cooling flow. We find no
strong evidence for multiphase gas in the cluster core except in the
20-40kpc region where the radio lobes have disturbed the gas most. We
limit the mass deposition rate within the central 100~kpc to
$12\Msunpyr$.  Only if the emission below about 1~keV is somehow
undetectable, such as occurs if there is intrinsic absorption, can
this rate be significantly increased to around $40\Msunpyr$.

We have quantified the heating rate and its distribution in the
cluster required to balance cooling if the gas is in a steady state.
However, the situation may not be steady. Any cooling flow is only
likely to occur and have become steady within the region where the
cooling time is a few billion year or less. The radio source is
clearly causing disruption to the south, east and west of the nucleus.
How much disruption is perhaps indicated by the abundance profile. Our
spectral fitting shows that the evidence for two temperatures is
almost entirely due to projection effects and
Fig.~\ref{rmjdeprojdataprof} gives the best description of the state
of the gas. The abundance shows a maximum at around 30 kpc. The
profile (which has a similar shape to that seen in the Centaurus
Cluster; Sanders \& Fabian 2002) is consistent with some mixing of gas
within, but not much beyond, that radius. Abundance gradients can be
due to metal injection by SN Ia in the central galaxy
\citep{DupkeWhite00,Ettori02}
and are expected to show a monotonic increase inward (unless the
metals are poorly mixed and cooling dominates; Morris \& Fabian 2002,
in preparation). A central decrease can be explained by some exchange
of gas between the centre and the region at 40--50~kpc (or further
out). Why intermediate radii have not been similarly affected is
puzzling.  Massive mixing should however have wiped out the gradient.
The high abundance of the gas at 10--40~kpc indicates that it has
probably resided close to the centre of the galaxy for an extended
period.

\section*{Acknowledgments}

We thank Stefano Ettori for the use of his software to make
exposure maps and to extract surface brightness profiles.
SWA and ACF acknowledge support by the Royal Society.

\bsp

\label{lastpage}

\end{document}